\documentclass[rsi,amsmath,amssymb,reprint]{revtex4-1}

\usepackage{graphicx}
\usepackage{dcolumn}
\usepackage{bm}

\usepackage[utf8]{inputenc}
\usepackage[T1]{fontenc}
\usepackage{mathptmx}

\begin{document}

\preprint{RSI/Shaposhnikov_1}

\title[Wide-Aperture Dense Plasma Fluxes Production Based on ECR Discharge in a Single Solenoid Magnetic Field]{Wide-Aperture Dense Plasma Fluxes Production Based on ECR Discharge \\in a Single Solenoid Magnetic Field}

\author{V.A. Skalyga}
 \altaffiliation[Also at ]{Lobachevsky State University \\ of Nizhny Novgorod, 603155 Nizhny Novgorod, Russia.}
\author{S.V. Golubev}
\author{I.V. Izotov}
\author{R.A. Shaposhnikov}
 \email[]{shaposhnikov-roma@mail.ru}
\author{S.V. Razin}
\author{A.V. Sidorov}
 \altaffiliation[Also at ]{Lobachevsky State University \\ of Nizhny Novgorod, 603155 Nizhny Novgorod, Russia.}
\author{A.F. Bokhanov}
\author{M.Yu.~Kazakov}
\author{R.L. Lapin}
\author{S.S. Vybin}
\affiliation{ 
Institute of Applied Physics, Russian Academy of Sciences, 603950 Nizhny Novgorod, Russia}

\date{\today}

\begin{abstract}
Results of experimental investigation of the ECR discharge in a single coil magnetic field as an alternative to RF and helicon discharges for wide-aperture dense plasma fluxes production are presented. A possibility of obtaining wide-aperture high density hydrogen plasma fluxes with homogenous transverse distribution was demonstrated in such a system. The prospects of using this system for obtaining high current ion beams are discussed.
\end{abstract}

\maketitle 

\section{Introduction}

One of the most widespread methods of plasma heating in systems of nuclear fusion is a high-energy neutral beam injection \cite{grisham}. Neutral beam injectors development is a complex and a multistage task. At the first step it is necessary to create a high-density plasma flux for the subsequent ion beam extraction. The beam current has to reach values of tens of Amperes to satisfy requirements of modern fusion facilities, and this could be realized only with using a large-scale multi-aperture extraction system, which implies a necessity of a spatially wide plasma fluxes production with a homogenous transverse distribution. Plasma sources, based on RF and helicon discharges, are mainly used for this purpose \cite{brown, shikhvotsev}.

An alternative approach based on the ECR discharge sustained by powerful millimetre wavelength gyrotron radiation was proposed in the Institute of Applied Physics of Russian Academy of Sciences (IAP RAS). The ECR discharge with a quasi-gasdynamic plasma confinement in a simple mirror trap was studied in the very first experiments \cite{golubev}. A possibility to obtain 100\%-ionized plasma with low content of molecular ions (at a level of several percent), density on the level of 10$^{13}$ -- 10$^{14}$~cm$^{-3}$, and electron temperature of 10 -- 100~eV was demonstrated \cite{skalyga_2014,skalyga_2016_1,skalyga_2017}. Combination of plasma parameters mentioned above seems to be optimal for light ion production. The ECR discharge has advantages of lower operating pressure, consequently, higher ionization degree, and the electron temperature being closer to the optimal one for production of pure atomic beams when compared to RF discharges. Additionally, plasma heating with microwaves allows to use quasi-optical methods of its injection into a discharge, which eliminates the need of biasing the microwave generator, thus reducing electric power consumption at a high-voltage platform of an ion source. This factor favourably distinguishes the ECR discharge from other plasma production methods. The novelty of the present work is the use of a single solenoid field (magnetic field is produced by the only magnetic coil) instead of a simple mirror trap, when compared to what has been done at IAP RAS recently. Unlike simple mirror traps, the longitudinal plasma confinement in a single coil field is provided only by the ambipolar potential distribution, which accelerates ions and slows down electrons resulting in lower plasma lifetime. One could also expect lower ionization degree in this case, however it was demonstrated that energy input provided by powerful gyrotron is enough for sufficient hydrogen dissociation and ionization even under conditions of a poor confinement. The numerical simulations based on the theoretical model of plasma expanding along the magnetic field lines to a metal wall \cite{abramov} demonstrates that existing gyrotron systems are capable of providing the necessary heating power for a large-scale plasma production and dense proton fluxes formation with the parameters required in modern neutral injection systems. Also such a simple magnetic configuration is convenient for further scaling and looks technically and technologically attractive.

\section{Experimental facility}

\begin{figure*}
\includegraphics[width=1\linewidth]{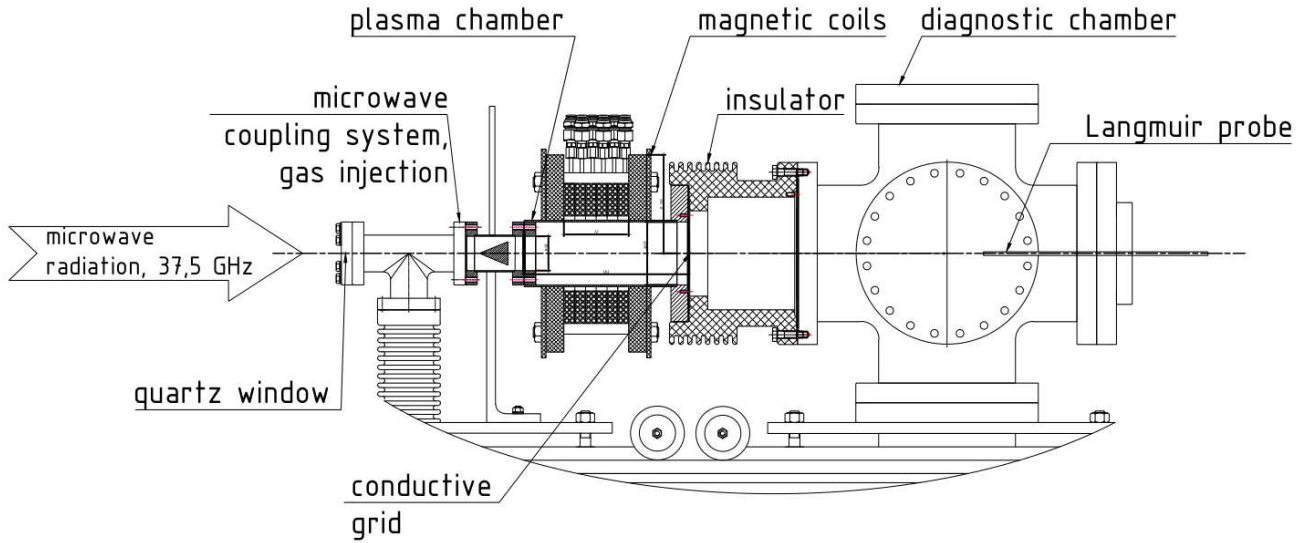}
\caption{\label{fig:scheme} Experimental facility scheme (gyrotron is not shown).}
\end{figure*}

Experiments were conducted at the IAP RAS using SMIS 37 facility \cite{skalyga_2014,skalyga_2016_1,skalyga_2017} (Fig.\ref{fig:scheme}), modified for described experiments. The gyrotron radiation with frequency 37.5~GHz, power up to 100~kW and pulse duration of 1.5~ms was used for discharge ignition and plasma heating. Electromagnetic radiation was focused by a special quasi-optical system to the centre of the discharge chamber with a diameter of 68~mm and a length of 250~mm placed inside a pulsed magnetic coil. 

The wedge-like coupling system was located inside the discharge chamber in order to prevent plasma flux reaching the quartz vacuum window. The microwave injection part has an inner diameter of 38~mm, whereas the plasma chamber -- 68~mm (see Fig.\ref{fig:scheme}). An influence of that fact on obtained result is discussed in the conclusion section of the present paper. A metal grid with a transparency of 70\% was installed at the farther end of the discharge chamber, opposite to the injection side, in order to create a microwave cavity to improve the efficiency of plasma heating. In the presented research the metal grid was a part of the discharge chamber and, accordingly, they were under the same potential, namely, they were grounded. The magnetic field in the centre of the coil was varied from 1 to 3~T, while the resonance field value for the radiation frequency of 37.5~GHz is 1.34~T.

\begin{figure}
\includegraphics[width=1\linewidth]{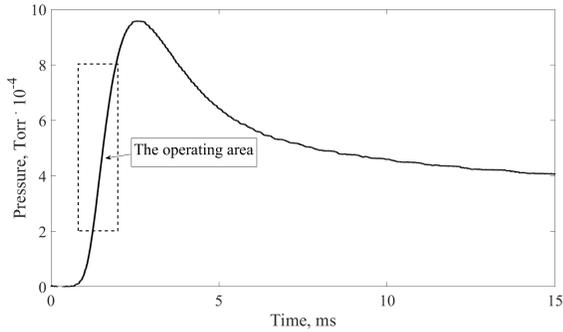}
\caption{\label{fig:pulse} Plasma chamber gas pressure pulse structure.}
\end{figure}

The neutral gas (hydrogen) was injected into the discharge chamber along the field axis through the gas supply line integrated into the microwave coupling system. The gas injection system was arranged as follows. A special tank filled by the neutral gas was connected by a tube to a pulsed valve placed on the discharge chamber. Pressure in the discharge chamber was determined by two parameters: gas pressure above the valve and time delay between the moment of the valve opening and the beginning of the microwave pulse. Pressure measurements in the discharge chamber were conducted using a gas discharge lamp before the experiments start in order to calibrate the gas feeding system. Thus, during the experiments the gas pressure in the discharge chamber was controlled by varying of that set of the gas inlet system parameters. The gas pressure was not measured in case of the discharge ignition. An example of gas pressure time dependence during the gas injection pulse is shown in Fig.\ref{fig:pulse}. The microwave pulse triggering time was tuned to ignite the plasma at the rising edge of the pressure whilst it was in the range of $2\times 10^{-4}$ -- $8\times 10^{-4}$~Torr. The residual gas pressure in the diagnostic chamber was on the order of $10^{-6}-10^{-7}$~Torr. 

The plasma flux was studied with a single flat Langmuir probe with 1~mm$^2$ square, placed on a 3D movable rod mounted on the back flange of the diagnostic chamber providing measurement both in longitudinal and transverse directions with respect to the magnetic field. The probe was biased negatively, thus measuring the ion saturation current.

A series of experiments on the optical spectroscopy of the ECR plasma were performed to determine the plasma density in the discharge. An MS5204i monochromator-spectrograph manufactured by SOL Instruments was used to register the plasma emission spectrum. The spectrograph was equipped with removable diffraction gratings (1800~grooves/mm, blaze wavelength 270~nm; 1200~grooves/mm, blaze wavelength~400 nm) and was capable of registering radiation in the range of 200 -- 900~nm. A CCD camera connected to a PC was used as a radiation detector having resolution of 14$\times$2048 pixels, pixel size -– 14$\times$14~$\mu$m. The instrumental function of the spectrograph was equal to 0.16~nm for an entrance slit width of 60~$\mu$m. Plasma emission in optical range going through the optical flange, specially mounted for these studies on the right end of the diagnostic chamber, was collected with a quartz lens into an entrance of a fiber, whose output was coupled with the entrance slit of the spectrograph. Thus, we analysed the plasma emission received along the system axis.

\section{Plasma density measurements}

As it was mentioned above, plasma density in the ECR discharge is one of the key parameters which determines the maximum value of extracted ion beam current, and it could be enhanced by means of increase of the frequency of microwaves heating the plasma. Similar to \cite{skalyga_2017} a series of experiments on spectroscopy were performed to determine the plasma density according to Stark broadening of the lines of Balmer series of atomic hydrogen \cite{griem}. Analysis of hydrogen lines of Balmer series (transition from highly excited electron levels to the level with principal quantum number $n=2$) was performed, as this series lies in visible and near-ultraviolet range. In case of 37.5~GHz ECR discharge, plasma density is expected to be on the level of $N_e=(1-5)\cdot10^{13}$ cm$^{-3}$. H$_\alpha$ and H$_\beta$ lines broadening is hardly detectable for a given plasma density. Therefore, analysis of shorter wavelength lines (H$_\gamma$ and H$_\delta$) is needed, as Stark splitting of electron level in hydrogen atom, which leads to a line broadening, is proportional to $n^2\cdot N_e^{3/2}$ ($n=3$ for H$_\gamma$, $n=6$ for H$_\delta$).

\begin{figure}[h!]
\includegraphics[width=1\linewidth]{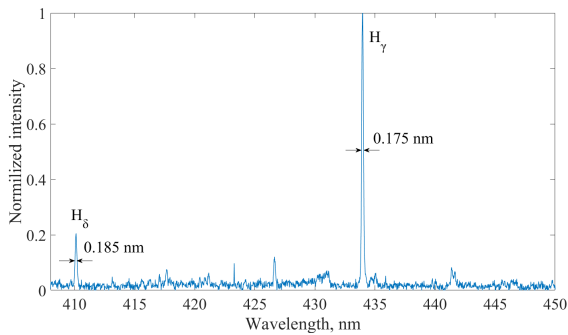}
\caption{\label{fig:spectrum} Plasma emission spectrum showing H$_\gamma$ and H$_\delta$ lines.}
\end{figure}

Fig.\ref{fig:spectrum} shows a part of plasma emission spectrum containing H$_\gamma$ (434.1~nm) and H$_\delta$ (410.1~nm) lines. These lines had a noticeable broadening, which differed from instrumental function significantly at certain experimental conditions. Obtained full width at half maximum (FWHM) were 0.175~nm, and 0.185~nm for H$_\gamma$ and H$_\delta$ lines respectively.  Real broadening was estimated as $\Delta \lambda$=($\Delta \lambda^2_{meas}$ - $\Delta \lambda^2_{inst}$)$^{0.5}$, where $\Delta \lambda_{meas}$~–-~ measured FWHM of a line, and $\Delta \lambda_{inst}$~-–~instrumental function equal to 0.16~nm. Stark broadening of these lines is the main effect, since other effects are negligible at our experimental conditions. For example, Doppler broadening $\Delta \lambda/\lambda$ is on the order of 10$^{-5}$ even at gas temperature of 2000~K (see \cite{ochkin}), whereas observed broadening is on the level of 10$^{-4}$ and could be induced only by Stark effect. Calculations and various experiments \cite{ochkin} show that for hydrogen atom in case of Stark effect lines full widths at half maximum (FWHM) depends on plasma density as $\Delta \lambda=C\cdot N_e^{2/3}$, where C is a constant slightly depending on electron density and temperature (C varies only 20 -- 30\% in 5000 -– 20000~K temperature range and in 10$^{14}$ -– 10$^{17}$~cm$^{-3}$ density range). 
 
This method allows to estimate the plasma density accurate to the order of magnitude that is why it was difficult to establish dependencies of plasma density on system parameters. In performed experiments plasma density according to this data varied in the range of 10$^{13}$ -– 10$^{14}$~cm$^{-3}$. The obtained values coincide in the order of magnitude with the expected ECR plasma density for gyrotron frequency used in experiments.

\section{Plasma flux transverse profile measurements}

Plasma flux transverse profile measurements were provided with a Langmuir probe mounted on 3D movable rod as it shown in Fig.\ref{fig:scheme}, allowed to obtain flux distribution at various distances from the centre of the discharge. The voltage U = -40~V corresponding to the ion saturation current was applied to the probe to collect a signal proportional to the plasma flux density. 

\begin{figure}[h!]
\includegraphics[width=1\linewidth]{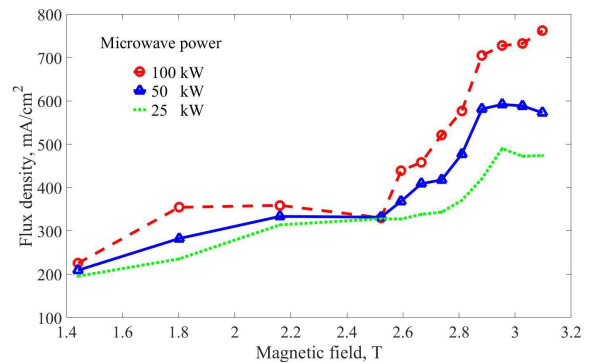}
\caption{\label{fig:density} Plasma flux density dependence on the magnetic field value at the coil centre. Gas pressure is $3\times 10^{-4}$~Torr.}
\end{figure}

To study the discharge properties under various experimental conditions plasma flux density was measured for different values of microwave heating power and magnetic field at the system axis close to the metallic grid separating plasma and diagnostic chambers, i. e. 7.4~cm downstream the magnetic field maximum. These results obtained with a gas pressure equal to $3\times 10^{-4}$~Torr (measured at the leading edge of the microwave pulse) are shown in Fig.\ref{fig:density}.

It is clearly seen that the plasma flux density could reach 750~mA/cm$^2$ level. At the same time it should be mentioned that dependence on the magnetic field looks quite complicated and probably it could be explained by significant change in microwave-to plasma coupling efficiency together with the ECR surface shift along the plasma chamber. The effect of the ECR surface position would be studied in further experiments.

\begin{figure}[h!]
\includegraphics[width=1\linewidth]{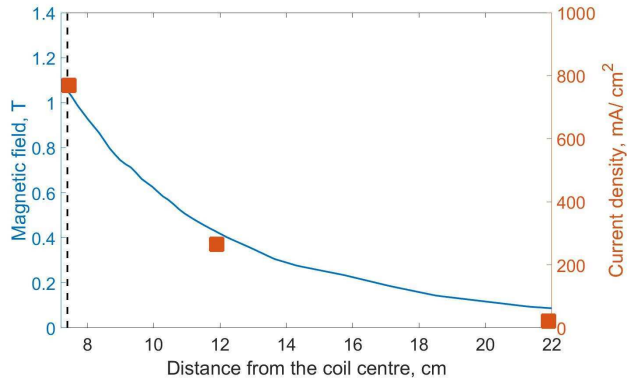}
\caption{\label{fig:mfield} The simulated magnetic field and the plasma flux density dependence from the longitudinal coordinate counted from the magnetic coil center. The grid position is indicated by the black dashed line.}
\end{figure}

The dependence of the flux density on the longitudinal coordinate was studied, performing measurements at a several distances from the coil centre at experimental conditions corresponding to the maximum flux density. The measurements are shown in Fig.\ref{fig:mfield} together with the magnitude of the magnetic field. It may be argued from the Fig.\ref{fig:mfield} that the plasma flux follows the magnetic field lines, which means that for real application plasma flux profile could be efficiently shaped by magnetic field distribution in expanding region.

The data shown in the Fig.\ref{fig:mfield} could be additionally used for as one more way of plasma density estimation. As far as the plasma flux density varies in proportion with the magnetic field value, it could be recalculation into the coil center, and it corresponds to the value of 2~A/cm$^2$. We can estimate the plasma density assuming that the flux has a velocity close to the ion sound and the electron temperature in the discharge 10 -- 30~eV, typical for SMIS 37 conditions. Such approach gives us the plasma density value of $2-4\times10^{12}$ cm$^{-3}$ and that corresponds to the ionization degree on the level of 50\%. These numbers are in a good agreement with the data presented above.

\begin{figure}[h!]
\includegraphics[width=1\linewidth]{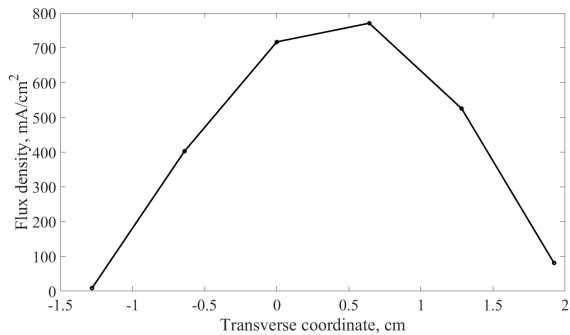}
\caption{\label{fig:distr} Plasma flux transverse profile at 7.5~cm from the coil center. Microwave power is 100~kW, gas pressure is $3\times10^{-4}$~Torr, magnetic field in the coil center is 3.1~T.}
\end{figure}

To demonstrate the homogeneity of the plasma flux, the Fig.\ref{fig:distr} shows its transversal distribution in the diagnostic chamber close to the grid (see Fig.\ref{fig:scheme}). While the maximum density was 750~mA/cm$^2$, the corresponding total ion current reached 5~A, evaluated as $\int j(x)\cdot 2\pi xdx$, where j(x) is transverse flux density profile, x is transverse coordinate. Also it could be seen that plasma flux FWHM is equal to 3~cm, while the plasma chamber is 68~mm in diameter. 

\begin{figure}[h!]
\includegraphics[width=1\linewidth]{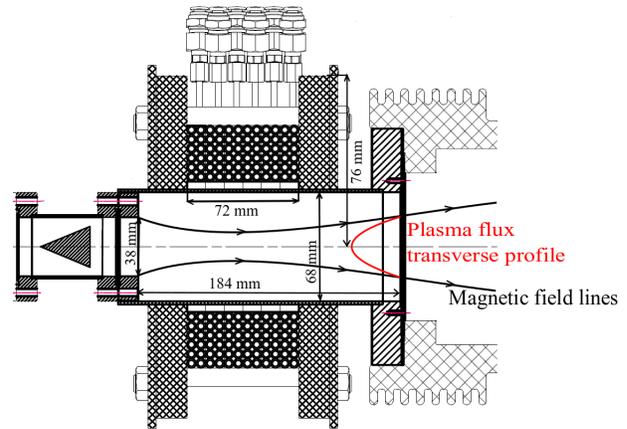}
\caption{\label{fig:mlines} The plasma chamber cross-section and magnetic field lines, presumably constraining the plasma flux.}
\end{figure}

That may be explained by the structure of magnetic field lines, which are shown in Fig.\ref{fig:mlines} together with the plasma chamber geometry. It is seen in Fig.\ref{fig:mlines} that the plasma flux is likely constrained by the diameter of microwave coupling system. This fact additionally underlines the importance of microwave injection scheme for the discharge parameters optimization. More studies on microwave coupling scheme are needed even in such a simple configuration.

\section{Conclusion}

Wide-aperture plasma fluxes from the ECR discharge in the magnetic field of a single solenoid were obtained. It was found that the flux density increases with the microwave power and magnetic field. Maximum value of flux density was j$_{max}$ = 750~mA/cm$^2$, and corresponding total ion current was equal to 5~A. It was also found that the plasma follows magnetic field lines while expanding into the diagnostic chamber. 

The obtained results may be supposedly improved with the increase of the frequency and the power of heating microwaves. An increase in the flux aperture could be provided by the use of discharge chambers of larger diameters simultaneously with a specific design of the microwave injection, optimised for more uniform electromagnetic field distribution in the volume. 

Obviously, a question about possibility of a continuous wave (CW) operation of such ion source arises. In our opinion, the CW operation could be realised after a number of the experimental facility upgrades. The first one is a change of the gyrotron system from pulsed to the CW one. This is not really a problem as far as a numerous number of such devices are produced commercially and widely used for plasma heating. It would be also necessary to provide a CW magnetic field, and that could be easily dome using permanent magnets. Probably the main difficulty on the way to the CW operation would be the plasma chamber and metallic grid cooling. The plasma chamber surface used in the described experiments has a square about 400~cm$^2$, thus taking into account that a regular water cooling could be used at power deposition level below 1~kW/cm$^2$, we can state that there is no any problem hear. The cooling of the metallic grid looks more complicated, but using of refractory materials (like molybdenum) which could stay under the temperatures high enough for efficient heat removal by radiation losses is a kind of usual technique. 

The obtained results demonstrate perspectives of the proposed system based on a single coil for wide-aperture plasma flux formation, including its probable application for neutral injectors development.

The further investigations will be dedicated to the obtaining of plasma fluxes with apertures of 100~cm$^2$ and flux density of 1~A/cm$^2$. We are also planning to optimize the gas inlet system. In the described experiments the gas inlet was carried out along the axis of the system. In this case the plasma flux had the density maximum on the axis. In the next experiments we are also going to use several tubes placed as a circle at a certain distance from the axis (a kind of a “ring” gas injection). This configuration could allow to provide a better control of transverse plasma flux profile and make it more homogenous. Another step in the future experiment would be devoted to a wide ion beam formation using this plasma source.

\begin{acknowledgments}
The reported study was funded by RFBR, project number 19-32-90079 and supported by Presidential Grants Foundation (Grant \# MD-2745.2019.2).
\end{acknowledgments}

\bibliography{Shap_single_sol}

\end{document}